\documentclass[prl,longbibliography,showpacs,twocolumn,superscriptaddress,amsmath,amssymb,verbatim]{revtex4-1}
\usepackage{graphicx}
\usepackage{lmodern}
\usepackage{epstopdf}
\usepackage{dcolumn}
\usepackage{bm}
\usepackage{subfigure}
\usepackage[pdftex,colorlinks=true,citecolor=blue,linkcolor=blue,urlcolor=blue,bookmarks=true]{hyperref}
\usepackage[sort&compress]{natbib}
\usepackage{braket} 

\newcommand{\bro}{Zn-brochantite}
\usepackage{units}

\usepackage{caption}
\begin{document}

\preprint{APS/123-QED}

\title{Kondo screening in a charge-insulating spinon metal}

\author{M. Gomil\v sek}
\affiliation{Jo\v{z}ef Stefan Institute, Jamova c.~39, SI-1000 Ljubljana, Slovenia}
\affiliation{ Centre for Materials Physics, Durham University, South Road, Durham, DH1 3LE, UK}
\author{R. \v Zitko}
\affiliation{Jo\v{z}ef Stefan Institute, Jamova c.~39, SI-1000 Ljubljana, Slovenia}
\affiliation{Faculty of Mathematics and Physics, University of Ljubljana, Jadranska c.~19, SI-1000 Ljubljana, Slovenia}
\author{M. Klanj\v sek}
\affiliation{Jo\v{z}ef Stefan Institute, Jamova c.~39, SI-1000 Ljubljana, Slovenia}
\author{M. Pregelj}
\affiliation{Jo\v{z}ef Stefan Institute, Jamova c.~39, SI-1000 Ljubljana, Slovenia}
\author{C. Baines}
\affiliation{Laboratory for Muon Spin Spectroscopy, Paul Scherrer Institute, CH-5232 Villigen PSI, Switzerland}
\author{Y. Li}
\affiliation{Experimental Physics VI, Center for Electronic Correlations and Magnetism, University of Augsburg, 86159 Augsburg, Germany}
\author{Q. M. Zhang}
\affiliation{National Laboratory for Condensed Matter Physics and Institute of Physics, Chinese Academy of Sciences, Beijing 100190, China}
\affiliation{School of Physical Science and Technology, Lanzhou University, Lanzhou 730000, China}
\author{A. Zorko}
\email{andrej.zorko@ijs.si}
\affiliation{Jo\v{z}ef Stefan Institute, Jamova c.~39, SI-1000 Ljubljana, Slovenia}

\date{\today}

\maketitle

{\bf
The Kondo effect, an eminent manifestation of many-body physics in condensed matter, is traditionally explained as exchange scattering of conduction electrons on a spinful impurity in a metal \cite{kondo1964resistance,hewson1997kondo}.
The resulting screening of the impurity's local moment by the electron Fermi sea is characterized by a Kondo temperature $T_K$, below which the system enters a non-perturbative strongly-coupled regime. 
In recent years, this effect has found its realizations beyond the bulk-metal paradigm in many other itinerant-electron systems, such as quantum dots in semiconductor heterostructures \cite{goldhaber1998kondo, cronenwett1998tunable} and in nanomaterials \cite{nygaard2000kondo, park2002coulomb, yu2004kondo}, quantum point contacts \cite{cronenwett2002low, iqbal2013odd}, and graphene \cite{chen2011tunable}.
Here we report on the first experimental observation of the Kondo screening by chargeless quasiparticles.
This occurs in a charge-insulating quantum spin liquid, where spinon excitations forming a Fermi surface take the role of conduction electrons.
The observed impurity behaviour therefore bears a strong resemblance to the conventional case in a metal.   
The discovered spinon-based Kondo effect provides a prominent platform for characterising and possibly manipulating enigmatic host spin liquids.}

\begin{figure}[b]
\includegraphics[trim = 0mm 0mm 0mm 0mm, clip, width=1\linewidth]{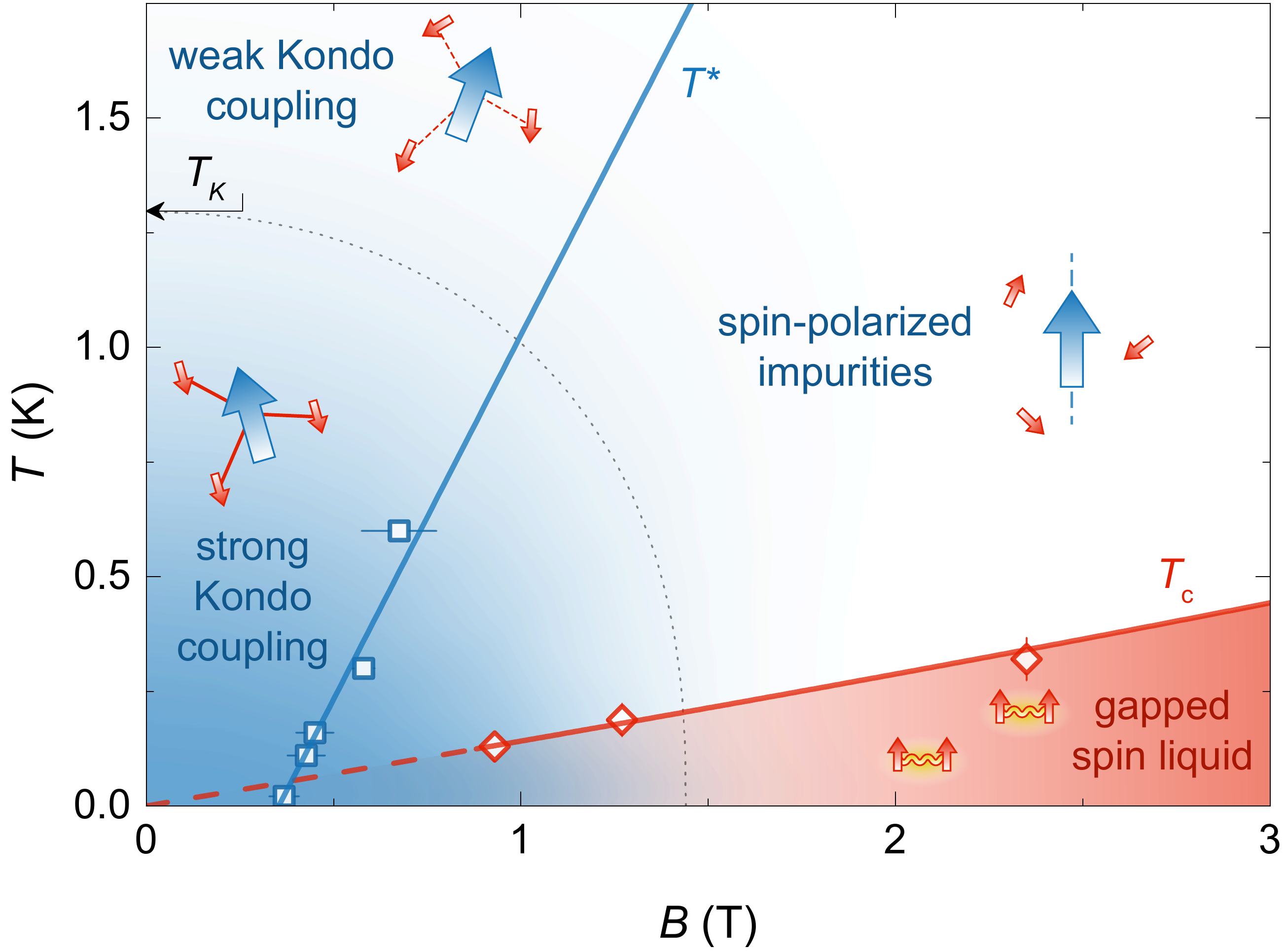}
\caption{
{\bf Phase diagram of \bro.}
Kondo screening of impurity spins (large arrows) by spinons (small arrows) in the spin-liquid state of \bro, with the Kondo temperature $T_K=1.3$~K.
The intensity of the blue hue indicates the slope of the imaginary part of the dynamical impurity spin susceptibility $\chi''(\omega)/\omega$ at frequency $\omega \rightarrow 0$.
The dotted line marks the contour corresponding to the value at $T_K$ and $B=0$. 
The temperature $T^*$ denotes the crossover between the low-field-Kondo regime and the field-polarized state after Kondo-resonance splitting (the field direction is indicated by the vertical dashed line). 
$T^*(B)$ yields the $g$ factor $g=2.3(3)$.
The spinon metal undergoes an intrinsic field-induced spinon-pairing transition into a gapped spin-liquid state below a field-dependent critical temperature $T_c$, where the density of spinon pairs increases with the applied field (red hue) \cite{gomilsek2017field}. 
This instability, which is analogous to the formation of Cooper pairs in superconductors, is characterized by an order of magnitude smaller effective $g$ factor, which clearly differentiates between the two effects.
}
\label{fig1}
\end{figure}

The Kondo screening of a magnetic impurity embedded in a quantum spin liquid (Fig.~\ref{fig1}), a highly-quantum-entangled yet magnetically disordered state where charge degrees of freedom are frozen, has been the subject of theoretical investigations for more than two decades \cite{khaliullin1995magnetic, sushkov2000spin, kolezhuk2006theory, florens2006kondo, ribeiro2011magnetic, dhochak2010magnetic, vojta2016kondo}.
In this case, the itinerant electrons of the standard Kondo picture
 are effectively replaced by emergent fractional magnetic excitations, inherent to any spin liquid.
Due to a large variety of essentially different low-energy excitations, spin liquids should be particularly versatile hosts of Kondo physics. 
In the case of spinon excitations with a Fermi surface --- a spinon metal --- a Kondo-like effect similar to the one in an ordinary metal is expected \cite{kolezhuk2006theory,ribeiro2011magnetic}.
However, emergent gauge fields mediating spinon--spinon interactions could cause deviations from the generic Fermi-liquid Kondo behaviour.
As the spin-liquid Kondo effect has not yet been conclusively confirmed by experiment, these theoretical predictions remain to be verified.  
\begin{figure*}[t]
\includegraphics[trim = 0mm 0mm 0mm 0mm, clip, width=0.9\linewidth]{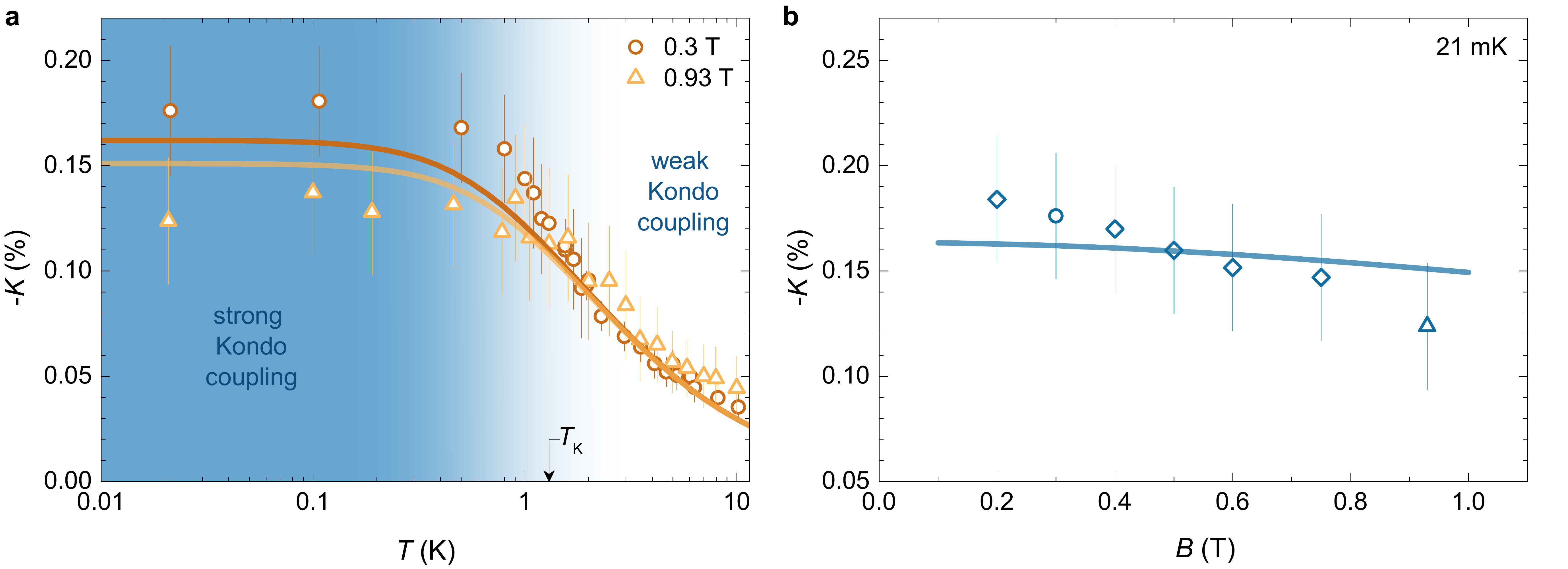}
\caption{
{\bf Static $\mu$SR signature of the Kondo effect.}
{\bf a}, The temperature dependence of the muon Knight shift $K$ (symbols) at two transverse magnetic fields in both the strong-Kondo-coupling (blue region) and the weak-Kondo-coupling regimes.
The dataset at 0.3 T corresponds to previously published measurements \cite{gomilsek2016muSR} with a refined analysis (see Methods).  
The lines show NRG theoretical predictions for the Kondo effect of impurities with concentration $p=11(2)\%$ coupled to non-interacting spin-1/2 fermions with the magnetic coupling of $a=31\;\text{mT}/\mu_B$ and the Kondo temperature $T_{K}=1.3~{\text K}$. 
The low-temperature plateau in $K$ corresponds to a strongly reduced average impurity magnetic moment, $M=KB/a=0.017\mu_B$ at $B=0.3$~T, due to Kondo screening.  
{\bf b}, The field dependence of the experimental Knight shift (symbols) deep in the strong-Kondo coupling regime and the NRG theoretical prediction (line).}
\label{fig2}
\end{figure*}

\bro, ZnCu$_3$(OH)$_6$SO$_4$, is a particularly well-suited compound for investigating the Kondo effect in a spin liquid.
This quantum kagome antiferromagnet \cite{li2014gapless} is one of only a few  examples where a spin-liquid state remains stable down to the lowest experimentally accessible temperatures \cite{norman2016herbertsmithite}, three orders of magnitude below the dominant exchange interactions  \cite{gomilsek2016instabilities}.
As revealed in our recent muon spin rotation and relaxation ($\mu$SR) study \cite{gomilsek2016muSR}, the corresponding spinon excitations are gapless and form a spinon Fermi surface.
This investigation has further disclosed that impurities, originating from intrinsic 6--9\% Cu--Zn intersite disorder \cite{li2014gapless}, are strongly magnetically coupled to the spin liquid and that their magnetic behaviour is characterized by a low-temperature magnetization plateau with a strongly reduced magnetic moment. This could be a sign of Kondo physics.

In-depth $\mu$SR experiments on \bro~that are reported here indeed confirm the first realization of the Kondo effect in a charge insulator (Fig.~\ref{fig1}).
The impurity magnetization extracted from the muon Knight shift clearly demonstrates the crossover from the weak-Kondo-coupling regime at high temperatures to the strong-Kondo-coupling regime at low temperatures.
Furthermore, magnetic-field-induced changes at the Fermi level characteristic of the Kondo effect are detected through muon spin relaxation.
These experimental findings are supported by numerical renormalization group (NRG) calculations, which furthermore show that the Kondo screening might be affected by spinon--spinon interactions mediated by emergent gauge fields.

First, we prove that impurities in {\bro} are not free.
To show this, we use muons, which represent ideal local probes of impurity magnetism.
They are dominantly coupled to impurity spins at low temperatures, as demonstrated by a linear scaling of the muon-detected local susceptibility with impurity susceptibility \cite{gomilsek2016muSR}, yielding a magnetic coupling of $a=31\;\text{mT}/\mu_B$ between the muon spin and the surrounding impurities (see Methods).
The muon Knight shift $K$ (see Methods), which is proportional to the local impurity magnetization $M$ normalized by the applied magnetic field $B$,  $K=a M/ B$ \cite{yaouanc2011muon} 
increases monotonically with decreasing temperature (Fig.~\ref{fig2}a).
In the applied field of 0.3~T it reaches a plateau value of 0.18\% at the lowest temperatures .
This yields an average magnetic moment of $M=KB/a=0.017\mu_B$ per magnetic Cu$^{2+}$ ion.
As the spin-1/2 impurity concentration in \bro~is estimated at 6--9\% \cite{li2014gapless}, the average magnetic moment per impurity is in the range of 0.19--0.28~$\mu_B$. 
This is significantly short of the full value of 1.0--1.2~$\mu_B$ expected for the Cu$^{2+}$ ions, the value that would be observed for low-temperature saturation of free impurity spins in an applied magnetic field.
Furthermore, the low-temperature Knight-shift saturation plateau gets reduced by only a factor of $\sim$1.4 in a more than three-times larger field of 0.93~T (Fig.~\ref{fig2}a), not by the factor $0.93/0.3=3.1$ that would apply to free impurities.

As an alternative to free impurities, one should consider an interpretation based on impurity--impurity coupling.
A reduction of the average impurity magnetic moment could naturally result from the formation of spin singlets in even-sized clusters of, e.g., antiferromagnetically coupled impurities.
However, to reproduce the experimental reduction of the average impurity moment, a large majority of impurities would have to be clustered, which is not compatible with their moderate concentration of 6--9\% (see Supplementary Fig.~1) assuming couplings only between nearest-neighbour randomly-positioned impurities  (see Supplementary Information).
If longer-range couplings between impurities are present --- either due to random far-neighbour exchange couplings, long-range dipolar couplings, or effective RKKY couplings mediated by the spinon Fermi sea --- a random-singlet phase should form \cite{kimchi2017valence} due to random distribution of impurities across the material.
If this were the case, a random-coupling spin-glass type of behaviour would be expected at the lowest temperatures \cite{kimchi2017valence}, which, however, contradicts the experimental results in several significant ways.
Specifically, in \bro~the temperature dependence of the magnetization (Fig.~\ref{fig2}a), the specific heat \cite{li2014gapless} and the muon-spin-relaxation rate \cite{gomilsek2016instabilities} are all smooth and do not feature any kink-like maxima characteristic of spin glasses \cite{mydosh2015spin}.

As the reduction of the impurity magnetization cannot be explained in an impurity-only picture, we next turn to the scenario of impurities coupled to the spin-liquid state. 
Namely, we consider the Kondo effect, where the impurity spin gets screened by spinful itinerant quasiparticles \cite{hewson1997kondo}.
In this case, the crossover between the high-temperature weak-coupling regime and the low-temperature strong-coupling regime is predicted to yield a smooth rise of impurity magnetization. 
Furthermore, a magnetization plateau corresponding to reduced impurity magnetic moment is expected at low temperatures \cite{zitko2011magnetization}.
This is precisely what is observed in \bro~(Fig.~\ref{fig2}a).
In order to quantitatively compare our experiment with the predicted Kondo impurity magnetization, we have performed NRG calculations within the framework of a non-interacting Fermi liquid (see Methods), as this method has proven indispensable in describing the Kondo effect in a fully general setting \cite{hewson1997kondo,wilson1975renormalization}.  
Taking into account the impurity $g$ factor $g=2.3$ (see below) and the magnetic coupling $a$, we achieve a good agreement between the experimental data and NRG predictions for both fields at all temperatures (Fig.~\ref{fig2}a). 
In this calculation, there were only two free fitting parameters: the impurity concentration and the Kondo temperature. 
The former turns out to be $p=11(2)\%$, which is consistent with the 6--9\% concentration estimate from various bulk measurements \cite{li2014gapless}.
For the Kondo temperature, on the other hand, we get $T_K=1.3(1)$~K, which is almost the same as the impurity Curie-Weiss temperature $|\theta|=1.2$~K determined from bulk susceptibility above $\sim$1~K \cite{li2014gapless}. 
This is in agreement with Kondo theory, which predicts a Curie-Weiss-like dependence of the impurity magnetization with $|\theta|\simeq T_{K}$ above $T_{K}$ \cite{hewson1997kondo}.

\begin{figure*}[t]
\includegraphics[trim = 0mm 0mm 0mm 0mm, clip, width=0.9\linewidth]{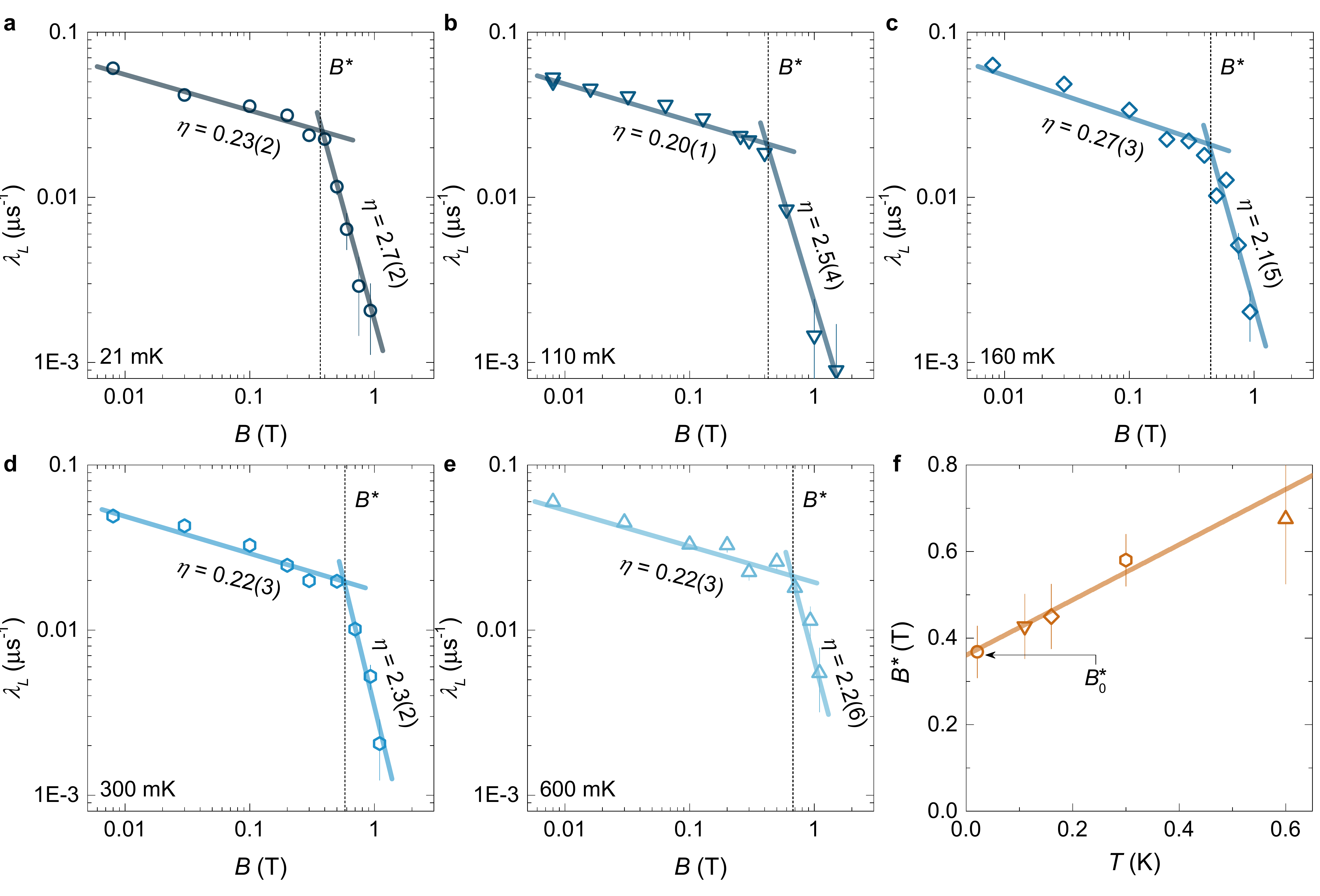}
\caption{
{\bf Dynamical $\mu$SR signature of the Kondo effect.}
{\bf a-e}, The field dependence of the muon spin relaxation rate $\lambda$ at various temperatures in the strong-Kondo-coupling regime (symbols).
Solid lines correspond to the power law $\lambda\propto B^{-\eta}$, while the dashed vertical lines indicate the anomaly field $B^*$, where $\lambda(B)$ changes its power-law dependence. This field corresponds to the Kondo-resonance splitting, a hallmark of the Kondo effect.
{\bf f}, The temperature dependence of the anomaly field $B^*$ (symbols). 
The line demonstrates the linear scaling $B^*=B^*_0+k_{B}T/(g\mu_{B})$, with the zero-temperature anomaly field $B^*_0=0.36(6)$~T and $g$ factor $g=2.3(3)$.}
\label{fig3}
\end{figure*}

A definitive proof of the spinon-based Kondo screening in {\bro} is the observation of Kondo-resonance splitting (Fig.~\ref{fig3}), a hallmark of the Kondo effect that is well known from the standard description of the strongly-coupled Kondo regime \cite{hewson1997kondo, costi2000kondo}.
The Kondo resonance is a characteristic enhancement of the local density of states (DOS) around the Fermi level associated with the coupling between the impurity spin and the itinerant host quasiparticles.
Its width at zero temperature is given by the Kondo temperature and increases with increasing temperature  \cite{costi2000kondo} due to thermal excitations.
Kondo-resonance splitting occurs above a finite critical magnetic field $B^*(T)$, when the field perturbs the resonance to such a degree that the splitting of its spin-up and spin-down components becomes larger than the zero-field resonance width.
Consequently, the DOS at the Fermi level gets decreased above $B^*$.
The muon spin relaxation rate $\lambda$ should also decrease above the same field.
This is because $\lambda$ probes local-magnetic-field fluctuations \cite{yaouanc2011muon} and thus couples to magnetic excitations at the Fermi level in a spinon metal, in close resemblance to nuclear magnetic resonance  (NMR) spin-lattice relaxation rate in a standard metal, which is proportional to the squared DOS at the Fermi level \cite{abragam1961principles}.
Indeed, we have already reported an anomalous reduction of the muon spin relaxation rate $\lambda$ above $B^*\sim 0.4$~T \cite{gomilsek2016muSR}.
However, as those data were limited to a single temperature of 110~mK, only a conjecture of a possible field-induced instability of the spin liquid could be made at that time. 

Our new, more comprehensive field-dependent muon spin relaxation-rate data collected at various temperatures below $T_K$ (Fig.~\ref{fig3}) demonstrate instead that the anomaly at $B^*$ is a manifestation of the Kondo resonance splitting.
We find that the anomaly field $B^*$ scales linearly with temperature, $B^*=B^*_0+\frac{k_{B}T}{g\mu_{B}}$, where $k_{B}$ and $\mu_{B}$ are the Boltzman constant and the Bohr magneton, respectively, and $B^*_0=0.36(6)$~T is the extrapolated zero-temperature anomaly field.
The extracted $g$ factor $g=2.3(3)$ is the same as the value $g\sim2.2$ found in the low-temperature Schottky impurity contribution to specific heat \cite{li2014gapless} and is typical for the Cu$^{2+}$ moments in an oxygen crystal-field environment.
Importantly, this $g$ factor is an order of magnitude larger than the effective $g$ factor that corresponds to an NMR-observed intrinsic field-induced instability of the spinon-metal state in {\bro} below a field-dependent critical temperature $T_c$ (Fig.~\ref{fig1}) due to spinon pairing \cite{gomilsek2017field}.
Therefore, the muon-detected anomaly at $B^*$ is not related to the latter instability.
On the other hand, in the Kondo picture, Kondo resonance splitting should occur at zero temperature at the field of the order of magnitude given by $\frac{k_{B}T_{K}}{g\mu_{B}}=0.8$~T \cite{costi2000kondo,kretinin2011spin}, which agrees reasonably well with experiment.
The fact that $B^*_0$ is finite is a distinctive feature of the Kondo effect.
Furthermore, $B^*$ is predicted to increase linearly with temperature due to an increasing width of the Kondo resonance \cite{costi2000kondo}, again in agreement with experiment (Fig.~\ref{fig3}f). 
Finally, we note that the power-law dependence $\lambda\propto B^{-\eta}$ changes at $B^*$ from being rather weak ($\eta =0.23(3)$) at $B<B^*$ to being much stronger ($\eta =2.4(3)$) at $B>B^*$ (Fig.~\ref{fig3}a-e).
The high-field power is close to 2, the value characteristic of a single-exponential time decay of the local-field autocorrelation function \cite{yaouanc2011muon} that is consistent with effectively quasi-free impurities above the Kondo resonance splitting field.
The low-field power $\eta =0.23(3)$, on the other hand, is close to 0, the value expected for a spinful impurity in a metal in the low-field strong-Kondo-coupling regime \cite{hewson1997kondo}.

The power-law muon relaxation $\lambda\propto B^{-0.2}$ (Fig.~\ref{fig3}a-e) at low fields instead of the expected field independence and the apparently systematically faster decrease of the impurity magnetization with increasing magnetic field than the simplest non-interacting Kondo theory predicts for the strong-Kondo-coupling regime (Fig.~\ref{fig2}b) are signs of a modified Kondo response in {\bro} that can be due to spinon--spinon interactions.
From the theoretical point of view, it was predicted by Ribeiro and Lee \cite{ribeiro2011magnetic} that in a spin-1/2 U(1) spinon metal the local susceptibility at the impurity site is insensitive to gauge degrees of freedom, which mediate the spinon--spinon interactions.
However, this conjecture is based on retaining only the leading order in a large-$N$ expansion, which, in the case of the $U(1)$ spinon metal, corresponds to a random-phase approximation.
Therefore higher-order corrections could lead to deviations brought about by spin-liquid gauge fields \cite{ribeiro2011magnetic}.

We find that the experimentally implied disagreements with the generic Kondo impurity response are indeed most likely to originate from emergent gauge fields. 
These deviations could, \textit{a priori}, be due to various perturbations to the conventional Kondo model of a rotationally-symmetric spin-1/2 magnetic impurity coupled to a bath of spinful non-interacting fermions with a constant density of states.
The perturbations of interest are those that break the full Kondo universality.
We have tested several of them by more elaborate NRG calculations of the impurity magnetization, as detailed in Methods.
We find that neither a finite spinon bandwidth, a finite $g$ factor of the spinons, a non-constant spinon DOS, an anisotropy of the Kondo coupling, nor a partial gap opening in the spinon DOS in an applied magnetic field can account for the experimental results (Supplementary Fig.~2).
Moreover, even a finite impurity concentration leading to a dilute Kondo lattice with coupling between the impurity spins cannot be responsible for the observed response (Supplementary Fig.~3).
The only plausible explanation of the possibly systematic discrepancy between the muon Knight shift at low temperatures and the Kondo theory that we can envisage is therefore based on non-negligible spinon--spinon interactions mediated through emergent spin-liquid gauge fields.
As regards this experimental observation, the gauge fields phenomenologically manifest in an effective renormalization of the relevant $g$ factors (Supplementary Fig.~4), whatever their microscopic effect may be.
The importance of gauge-field fluctuations in {\bro} is also directly reflected in the intrinsic spinon-pairing instability of the spinon-metal state, as well as in a power-law dependence of the NMR relaxation rate $1/T_1\propto T^{0.8}$ in the spinon-metal regime \cite{gomilsek2017field}, instead of the Korringa relation $1/T_1 \propto T$ that holds for non-interacting fermions \cite{abragam1961principles}.
It is expected that gauge-field fluctuations should introduce corrections to dynamical spin correlation functions with a power-law dependence \cite{itou2010instability, metlitski2015cooper}.
However, more rigorous predictions are needed to accurately assess the importance of gauge fields in the spinon-based Kondo effect, even though such calculations for interacting itinerant particles currently represent a huge theoretical challenge.

As the predicted Kondo response greatly depends on the exact nature of the host spin liquid \cite{kolezhuk2006theory}, Kondo-coupled impurities could, firstly, be utilized to study spinon excitations and the corresponding gauge fields in this enigmatic state of matter. 
This is a promising alternative to direct experimental studies in local observables, which are very challenging  \cite{han2012fractionalized,paddison2016continuous,klanjsek2017high,jansa2018observation}, because spin liquids lack symmetry-breaking order and their spinon excitations carry no charge, making them notoriously difficult to detect.
A further advantage is that the impurity response to the external field is usually much larger than the intrinsic spin-liquid response, which is dominated by singlets.
Secondly, certain spin liquids feature topologically-protected quasiparticles, e.g., non-Abelian anyons \cite{kitaev2003fault,jansa2018observation}, which, in principle, makes them promising candidates for robust quantum computing \cite{kitaev2003fault}.
However, finding a way of manipulating these states in a solid-state device is a key outstanding issue.
The required level of control could potentially be achieved by exploiting the Kondo coupling, by manipulating impurities to achieve the desired change in the spin liquid. 
One concrete theoretical proposal in this spirit is based on the Kitaev honeycomb spin-liquid, where the weak- and strong-Kondo-coupling limits are predicted to correspond to different topologies of the spin-liquid ground state \cite{dhochak2010magnetic}.    
%


%
\section{Acknowledgements}
The authors acknowledge fruitful discussions with T.~Lancaster and D.~Manevski. 
This work is partially based on experiments performed at the Swiss Muon Source S$\mu$S, Paul Scherrer Institute, Villigen, Switzerland.
The financial support of the Slovenian Research Agency under programs No.~P1-0125 and No.~P1-0044 and project No.~J1-7259 is acknowledged.
M.G. is grateful to EPSRC (UK) for financial support (grant No. EP/N024028/1). 
Q.M.Z was supported by the Ministry of Science and Technology of China (2016YFA0300504 \& 2017YFA0302904) and the NSF of China (11774419 \& 11474357).

\section{Author contributions}
A.Z. conceived, designed and supervised the project. 
M.G. and A.Z. performed the $\mu$SR measurements, with technical assistance of C.B., and analysed the data.
R.\v{Z}. carried out the NRG calculations.
M.G. developed the percolation-theory-based model for impurity-cluster spin.
Y.L. and Q.M.Z. synthesized and characterized the sample.
All authors discussed the results.
A.Z. wrote the paper with feedback from all authors.
  
\section{Competing interests}
The authors declare no competing interests.
\section{Additional information}
Supplementary information is available for this paper.\\

Correspondence and requests for materials should be addressed to A.Z.
\newpage
\section{Methods}
\subsection{$\mu$SR measurements}
The $\mu$SR investigation was conducted on the LTF instrument at the Paul Scherrer Institute (PSI), Switzerland, on a $\sim$100\% deuterated powder sample from the same batch as the one used in our previous investigations \cite{gomilsek2017field, gomilsek2016instabilities, gomilsek2016muSR}.
The sample was glued onto a silver sample holder with diluted GE Varnish to ensure good thermal conductivity.
The measurements were performed between 21~mK and 10~K in various transverse (TF) and longitudinal (LF) applied fields $B\leqslant 1$~T with respect to the initial muon-spin polarization.

In the TF setup the initial muon polarization was tilted by $\sim$45$^\circ$ away from the beam/field direction. 
Its component perpendicular to the field was detected by a set of detectors.
The muon asymmetry $A(t)$, which is proportional to the muon polarization \cite{yaouanc2011muon}, was measured and modelled by
\begin{eqnarray}
\nonumber
A_{TF}(t)&=A_1\cos\left(2\pi\nu_1 t +\phi \right){\text e}^{-\lambda_1 t}\\ 
&+A_2\cos\left(2\pi\nu_2 t +\phi \right ){\text e}^{-\lambda_2 t},
\end{eqnarray}
where the total initial muon asymmetry was $A_1+A_2=0.199(12)$ and the ratio of the two components was $A_2/A_1=22\%$.
A typical TF muon asymmetry curve is shown in the Supplementary Fig.~5 together with the corresponding Fourier transform. 
The fitting parameters are given in the Supplementary Methods.
The signal $A_1$ was attributed to the muons stopping in the sample and the signal $A_2$ to the muons stopping in the diamagnetic silver sample holder.
Since the background was diamagnetic, it decayed in time much more slowly than the intrinsic signal ($\lambda_2 \ll \lambda_1$) and its temperature-independent oscillation frequency $\nu_2$ could be used as the reference Larmor frequency.
The Knight shift was calculated as
\begin{equation}
K=(\nu_1-\nu_2)/\nu_2.
\end{equation}
Combining the new $\mu$SR measurements with those previously published in Ref.~\onlinecite{gomilsek2016muSR} we could better constrain the \textit{a priori} unknown backgrounds in both experiments by comparing and simultaneously fitting old and new measurements at the same values of $T$ and $B$. 
After a careful self-consistent fit of all measured data, we found that the best estimates of low-field Knight-shift values were reduced by $\sim$15\% in relative terms when compared to the data published in Ref.~\onlinecite{gomilsek2016muSR}. 
Consequently, the magnetic coupling between the muon spin and the impurities, $a=31\;\text{mT}/\mu_B$, is reduced by the same amount compared to the previously published value. 
The new values should be taken as the definitive ones.

In the LF setup, which was used to measure the longitudinal muon relaxation rate $\lambda$, the initial muon-spin polarization was along the beam/field direction and the muon asymmetry was measured along the same direction.
The asymmetry was modelled with a stretched-exponential model
\begin{eqnarray}
A_{LF}(t)&=A_1{\text e}^{-(\lambda t)^\beta}+A_2,
\end{eqnarray}
where the total initial asymmetry was $A_1+A_2=0.239(4)$, with the same ratio $A_2/A_1=22\%$ as in the TF experiment.
The same stretching exponent $\beta = 0.86(5)$ was found as in our previous LF studies \cite{gomilsek2016instabilities,gomilsek2016muSR}. 
A typical set of LF muon asymmetry curves at a selected temperature is shown in the Supplementary Fig.~5c.

\subsection{NRG calculations}

The numerical renormalization group (NRG) method
\cite{wilson1975renormalization, zitko2009energy} was utilized to compute the
impurity magnetization as a function of magnetic field and
temperature. 
We used the discretisation parameter $\Lambda=2$,
averaged over two discretisation grids, and kept a high number of
states to ensure full convergence. 
The Wilson's thermodynamic
definition \cite{wilson1975renormalization} of the Kondo temperature
where $T_K \chi_\mathrm{imp}(T_K) = 0.07$ was applied. 
Here $\chi_\mathrm{imp}$ is
the impurity contribution to the total-system's magnetic
susceptibility.  The reference calculations were performed for a spin-1/2
magnetic impurity with an isotropic Kondo exchange coupling to
a bath of non-interacting spin-1/2 fermions with a constant density of
states in a wide-band limit.

The Hamiltonian of the reference model is
\begin{equation}
H=\sum_{k\sigma} \epsilon_k c^\dag_{k\sigma} c_{k\sigma}
+ J \mathbf{S} \cdot \mathbf{s}(\mathbf{r}=0)
+ g \mu_B S_z B
\end{equation}
where the operators $c_{k\sigma}$ describe itinerant particles
with momentum $k$, spin $\sigma$, and energy $\epsilon_k$,
$J$ is the Kondo exchange coupling, $\mathbf{S}$ is the quantum-mechanical
spin-$1/2$ operator of the impurity, $\mathbf{s}(\mathbf{r}=0)$ is the spin
density of the itinerant particles at the position of the impurity,
$g$ is the impurity $g$-factor, $\mu_B$ the Bohr magneton, and $B$
the magnetic field.
$T_K$ is given approximately by $T_K=D \sqrt{\rho J} \exp(-1/\rho J)$. 
Here $D$ is the half-bandwidth of the itinerant-particles band and $\rho=1/(2D)$ is the corresponding density of states.

The reference model describes the intermediate and high
temperature ranges very well (Fig.~\ref{fig2}a). 
At low temperatures we find a deviation, which is still within the experimental uncertainty (and hence perhaps not even statistically significant),
but quite systematic.
The slope of the calculated $M(B)/B$ curve
is appreciably smaller than the slope of the measured $K(B)$ data
points at 21~mK (Fig.~\ref{fig2}b).
Therefore, additional sets of calculations were performed for various perturbations.

\textit{Finite spinon bandwidth}:
We first relaxed the assumption of the wide-band limit $T_K \ll D$,
where $D$ is the half-bandwidth of the conduction band. 
We considered the Kondo impurity
model with a flat band with $\rho J=0.7$ and $T_K/D \approx 0.5$, while the reference model had $\rho J=0.15$ and $T_K/D=4.2\times 10^{-4}$.
For this rather extreme case, the agreement with experiment is slightly improved at low temperatures, but is worse at high temperatures (Supplementary Fig.~2a). 

\textit{Finite spinon $g$ factor}: In the wide-band limit, the Zeeman
term of the itinerant particles can be neglected, since the impurity
magnetic susceptibility $\chi_i \propto 1/T_K$ is much larger than the
Pauli susceptibility of the band, $\chi_b \propto 1/D$.  The case of a
finite Zeeman term in the conduction band, which was included in the
calculation as described in Ref.~\onlinecite{hock2013numerical}, with
the spinon $g$ factor equal to the impurity $g$ factor was therefore
considered for a still rather narrow band so that $T_K/D \approx 0.2$
(by taking $\rho J=0.5$). The Hamiltonian thus had an additional term
$H'=g_b \mu_B B \sum_k (1/2)
(c^\dag_{k\uparrow}c_{k\uparrow}-c^\dag_{k\downarrow}c_{k\downarrow})$.
For this model, the slope of $M/B$~vs.~$B$ at low temperatures is actually reduced compared to the reference model (Supplementary
Fig.~2b).

\textit{Non-constant spinon DOS}: Since in a narrow-band situation the
details of DOS in the continuum, $\rho(\epsilon)$, may play a larger
role, we investigated some possible modifications for $\rho(0) J=0.5$,
e.g. a large slope of DOS $\rho$ across the Fermi level. This was
found to have little effect on the results. In Supplementary Fig.~2c
we show the case of a triangular-shaped DOS with $\rho(\epsilon)=\rho(0)
(D-\epsilon)/D$ for a narrow-band calculation with $T_K/D \approx 0.2$. 
Furthermore, potential scattering on the impurity site, a
perturbation of the form $H'=V n(\mathbf{r}=0)$ where $V$ is the local
potential and $n(\mathbf{r})$ the density of itinerant quasiparticles
at the position of the impurity, was also found to play little role
(Supplementary Fig.~2d). Only strong singularities in $\rho$ close to
the Fermi level could potentially lead to significant effects.
 
\textit{Kondo-coupling anisotropy}: The Kondo exchange coupling term
of the Hamiltonian can be separated into the transverse and
longitudinal parts, $H_K=J_\perp (S_x s_x+S_y s_y) + J_\| S_z s_z$. We
studied both the limit of a dominant Ising Kondo exchange coupling
($\rho J_\|=1.5$ and $\rho J_\perp=0.075$, so that $J_\|/J_\perp=20$)
and a dominant transverse coupling ($\rho J_\|=0.05$ and $\rho
J_\perp=0.5$, so that $J_\|/J_\perp=0.1$), yielding a comparable Kondo
temperature $T_K/D \approx 0.04$.  Even for these unphysically large
anisotropy ratios only small changes of $M/B$ curves
were found (Supplementary Fig.~2e). This can be explained by the fact
that the renormalization group flow toward the strong-coupling fixed
point for the anisotropic Kondo model tends to restore the isotropy at
low energy scales \cite{hewson1997kondo}.

\textit{Partial gap opening in the spinon DOS}:
At low temperatures and sizeable fields the quantum spin liquid in {\bro} becomes partially gapped through a field-induced spinon-pairing instability \cite{gomilsek2017field}.
The fraction of residual
ungapped spinons is $f=0.3$ at $B=1$~T and decreases with increasing field.  The behaviour for $B<1$~T is not known, therefore a linear extrapolation to low fields with $f=1$ at $B=0$ was made.
As the simplest approximation to incorporate the partial gap
opening in the calculations we assumed a modification of the
conduction-band DOS proportional to the gapped
fraction $1-f$ of spinons. For this gapped fraction we considered the
case of a hard gap of width $\Delta$, and the case of a BCS-like redistribution of the spectral weight with $\Delta_\mathrm{BCS}=\Delta$.
The partial gapping of spinons leads to reduced Kondo screening at very low
temperatures due to a decreased spinon DOS at the Fermi level, hence the impurity magnetization is enhanced
at low temperatures (Supplementary Fig.~2f).
This effect increases in strength as the magnetic
field is increased, thus inverting the slope of the $K(B)$ curve at
low temperatures, which clearly contradicts the experiment.   
We note that extreme versions
of the DOS reduction were assumed with a full suppression in the energy range
$|\epsilon|<\Delta$. The actual gap is very likely to be softer.

\textit{Finite impurity concentration}:
The impurity concentration of 6--9\% in Zn-brochantite \cite{li2014gapless} is
sufficiently large to prompt the question of possible
impurity--impurity-coupling corrections to the dilute-limit results. 
The relevant model to describe this situation
is the Kondo lattice model (KLM) with a random distribution of spins.
The corresponding Hamiltonian consists of a tight-binding lattice of
non-interacting itinerant fermions with density $n$ ($n=1$ is a
half-filled band) and additional local moments ($S=1/2$ spins) coupled by the Kondo exchange coupling on a random subset
of sites with concentration $p$. 
We calculated $M(B,T)$ using the real-space dynamical mean-field theory (RDMFT)
with the NRG as the impurity solver.
This is a conceptually simple (but numerically expensive) way to extend
the single-impurity NRG results to the case of a finite impurity
concentration. 
The calculations were performed on a $17 \times
17$ lattice with electron occupancy $n=0.5$ (quarter-filled band), for
$\rho J=0.4$.
For low impurity concentrations the curves
are all very similar -- the curves for $p=1\%$ and $p=10\%$ actually overlap within
the numerical errors (Supplementary Fig.~3). 
At $p=20\%$, some deviations are observable in the
high-field range, while the low-field results are still little
affected.
We thus conclude that the 6--9\% impurity concentration in
Zn-brochantite can be safely considered as corresponding to the
``dilute limit'' where the magnetization can be computed using the
single-impurity model. 
This is also in line with heavy-fermion systems where the magnetic
susceptibility and other quantities scale with the concentration of
magnetic ions over a surprisingly large concentration range
\cite{lin1987}.
A magnetization plateau that is found in the dense-impurity limit at intermediate fields (Supplementary Fig.~3) is consistent with theory \cite{kusminskiy2008,golez2013} and corresponds to a transition from a paramagnetic state to a ferromagnetic half-metal state, indicating the existence of a well-defined partial gap.

\textit{Renormalization of $g$ factors}:
Finally, we mention two cases that lead to theoretical predictions in better agreement with experiment, but do so for parameter values that either differ from the experimentally established ones or are clearly unphysical. An
example of the former case is the assumption of the impurity $g$ factor much
in excess of $g=2.3$. In Supplementary Fig.~4a we show the
calculation with $g=6.5$, which are in near-perfect agreement with experiment at all temperatures and fields. An example of the latter case is the assumption
of a large negative value for the $g$ factor of the itinerant fermions. In
Supplementary Fig.~4b we show the calculations for $g_b=-3$, which fit the low temperature results almost perfectly, but, however, deviate from experiment at high temperatures.

\section{Data availability}
The data that support the findings of this study are available from the
corresponding author upon reasonable request.

\begin{widetext}
\vspace{19cm}
\begin{center}
{\large {\bf Supplementary information:\\
Kondo screening in a charge-insulating spinon metal}}\\
\vspace{0.5cm}
M. Gomil\v sek,$^{1,2}$ R. \v Zitko,$^{1,3}$, M. Klanj\v sek,$^{1}$ M. Pregelj,$^1$ C. Baines,$^{4}$ Y. Li,$^{5}$ Q. M. Zhang,$^{6, 7}$ and A. Zorko$^{1,*}$
\vspace{0.3cm}

{\it
$^1$Jo\v{z}ef Stefan Institute, Jamova c.~39, SI-1000 Ljubljana, Slovenia\\
$^2$Centre for Materials Physics, Durham University, South Road, Durham, DH1 3LE, UK\\
$^3$Faculty of Mathematics and Physics, University of Ljubljana, Jadranska c.~19, SI-1000 Ljubljana, Slovenia\\
$^4$Laboratory for Muon Spin Spectroscopy, Paul Scherrer Institute, CH-5232 Villigen PSI, Switzerland\\
$^5$Experimental Physics VI, Center for Electronic Correlations and Magnetism,\\ University of Augsburg, 86159 Augsburg, Germany\\
$^6$National Laboratory for Condensed Matter Physics and Institute of Physics,\\ Chinese Academy of Sciences, Beijing 100190, China\\
$^7$School of Physical Science and Technology, Lanzhou University, Lanzhou 730000, China\\
}

\end{center}
\end{widetext}

\section{Supplementary Notes}

\subsection{Clustering of Impurities}\label{clusters}

In this section we demonstrate that any spin model with only short-range two-spin couplings between randomly-placed impurities obeys a stringent lower bound on the average value of the spin per impurity. This exact lower bound depends only on the topology of the possible couplings, and not on the precise type or strength of these couplings, and is a simple consequence of the rules for quantum addition of spin. Furthermore, we show that this bound is experimentally violated in \bro, refuting such a simple scenario of impurity magnetism in this system.

\subsubsection{Exact lower bound on average impurity spin}

Assuming that possible impurity sites are occupied uniformly randomly with probability (impurity concentration) $p$, the impurities will, by pure chance, form coupled impurity clusters of various sizes. Due to quantum spin addition rules, even-sized impurity clusters can, in principle, have total spin zero (forming singlets), but odd-sized clusters of spin-$1/2$ impurities must retain a total spin of at least $1/2$, a reduction by a factor of $1/s$, where $s$ is the (odd) cluster size, compared to the case of $s$ uncoupled impurities where the maximal spin is $s/2$. Coupled clusters of odd size $s$ (including lone, free impurities as $s = 1$ clusters) can thus only reduce the average spin per constituent impurity by a factor of $1/s$, regardless of the type of coupling. A lower bound on the spin reduction factor $f(p) = \braket{S}/S$, by which the average spin per impurity $\braket{S}$ can be reduced from its full value of $S = 1/2$ by impurity--impurity couplings, is then given for the whole impurity lattice by
\begin{equation}
f(p) \geq m(p) = \sum_{s\;\mathrm{odd}} \frac{w_s(p)}{s} 
\tag{1}
\label{clustersEq1}
\end{equation}
where $w_s(p) \geq 0$ is the probability that a randomly-selected impurity is part of a cluster of size $s$.

The cluster-size probabilities $w_s(p)$ are studied in uncorrelated site percolation theory \cite{stauffer1994introduction} and depend only on the topological connectivity of the underlying lattice of impurity sites. 
Two impurity sites are considered connected (adjacent) when they would have a non-zero impurity--impurity coupling if they were both actually occupied by an impurity. Note that the probabilities $w_s(p)$ do not depend on the specific type or strength of the impurity-impurity couplings, just on their spatial distribution and the distance range over which these couplings connect the impurities.

The first term, $w_1(p)$, in $m(p) \geq w_1(p)$ describes the contribution of lone impurities with no occupied adjacent impurity sites, which s behave are thus just free spins and contribute fully to the total impurity moment. The fraction of lone impurities is given by $w_1(p) = (1-p)^z$ and depends only on the coordination number $z$ (the number of sites adjacent to a given site) of the impurity-site lattice, assuming that $z$ is the same for all impurity sites.

Generalizing Eq.~(\ref{clustersEq1}) to impurities with arbitrary bare spin $S\geq 1/2$, we find that the lower bound on the spin reduction factor $f(p) = \braket{S}/S$ is given by
\begin{equation}
f(p) \geq m_S(p) = \alpha_S \cdot m(p) + (1 - \alpha_S) \cdot w_1(p)
\tag{2}
\label{clustersEq2}
\end{equation}
where $m(p)$ is defined by Eq.~(\ref{clustersEq1}), and where $\alpha_S = 1/(2S)$ for half-integer $S$ while $\alpha_S = 0$ for integer $S$ (giving $m_S \equiv w_1$). This difference in $\alpha_S$ is due to the fact that odd-sized clusters of size $s \geq 3$ must have a spin of at least $1/2$ for half-integer $S$, while they can form spin-zero singlets for integer $S$. On the other hand, even clusters can always, in principle, form spin singlets, while lone ($s = 1$) impurities always carry full spin $S$.

\begin{figure*}[t]
\includegraphics[trim = 0mm 0mm 0mm 0mm, clip, width=0.53\linewidth]{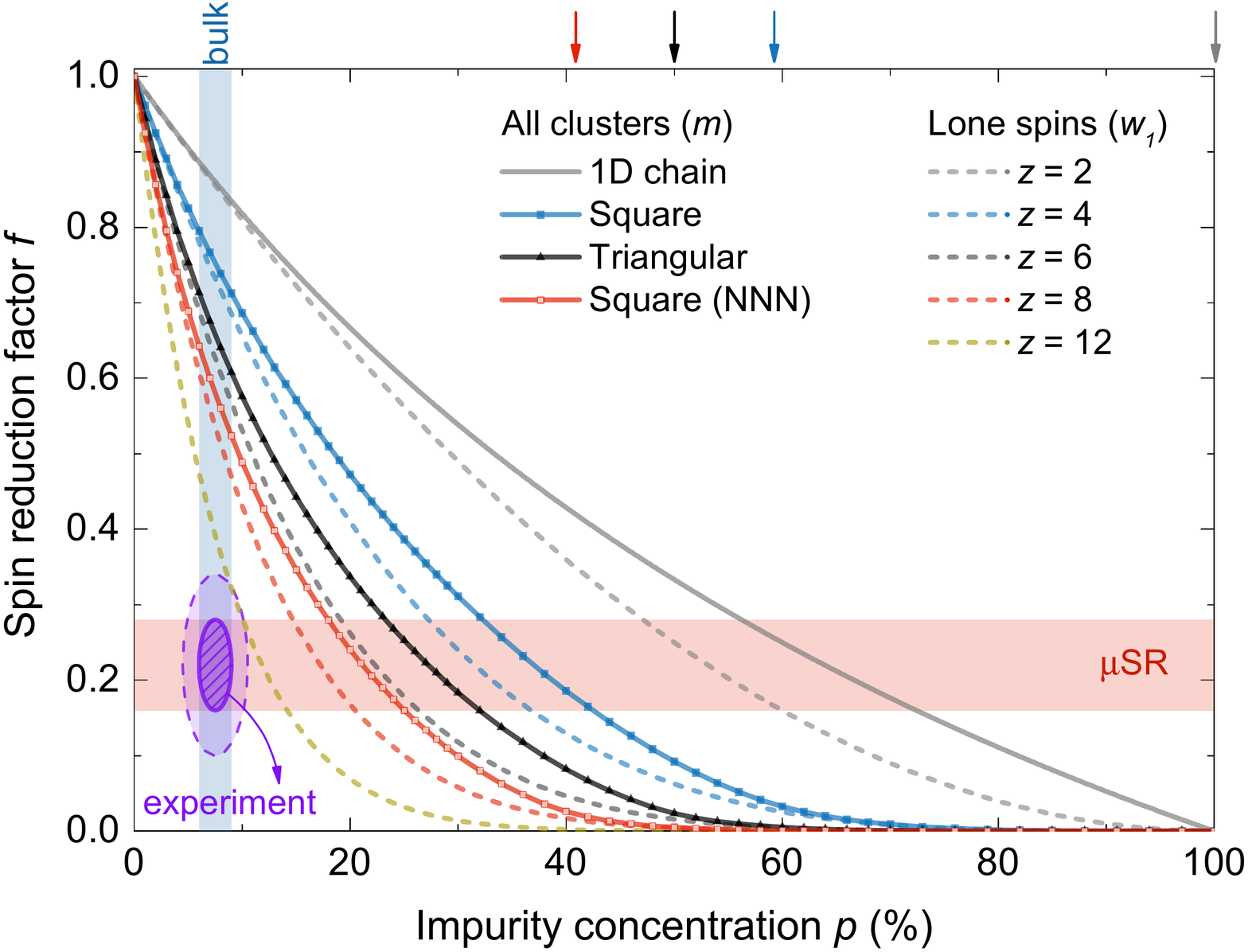}
\caption{
{\bf Supplementary Fig.~1 $\mid$ Spin-reduction factor due to impurity clusters.}
The solid lines are the lower bounds $m(p)$ on the average-spin reduction factor $f(p) = \braket{S}/S$ given by Eq.~(\ref{clustersEq1}) that were numerically calculated for various lattices and coordination numbers $z$. 
The coordination numbers of the lattices are: $z = 2$ for the 1D chain, $z = 4$ for the square lattice, $z = 6$ for the triangular lattice, and $z = 8$ for the square with additional next-nearest-neighbour (NNN) couplings.
The $z = 2$ 1D-chain case is analytically solvable with exact cluster probabilities $w_s(p) = s p^{s-1} (1-p)^2$ \cite{stauffer1994introduction} that, by using Eq.~(\ref{clustersEq1}), give an analytical result for the lower bound, $m(p) = (1-p)/(1+p)$. The dashed lines indicate the first term $w_1(p) = (1-p)^z$ of the lower bound $m(p) \geq w_1(p)$ and correspond to the contribution from lone impurity spins. The arrows indicate percolation thresholds $p_c$ for the corresponding lattices defined as the lowest impurity concentration above which the lattice contains an infinite spanning cluster with probability one \cite{stauffer1994introduction}. 
The shaded blue region indicates the 1-sigma (${\sim}68\%$ confidence-level) experimental interval for the impurity concentration $p$ \cite{li2014gapless}, while the shaded red region indicates the 1-sigma experimental interval for the spin reduction factor $f(p)$ in {\bro} (see the main text). The hashed violet  region is the combined 1-sigma region consistent with experiment, while the dashed violet ellipse encloses the 2-sigma (${\sim}95\%$ confidence-level) experimental region. 
All the theoretical lower bounds for the impurity spin lie well above the 2-sigma experimental region, refuting these as possible impurity-only explanations of our experimental results on Zn-brochantite.}
\label{clustersFig}
\end{figure*}

We stress that the lower bounds of Eq.~(\ref{clustersEq1}) and Eq.~(\ref{clustersEq2}) describe the lowest possible average impurity spin and derive from simple quantum spin addition rules in random impurity clusters. 
They and are thus independent of the specifics of the impurity--impurity couplings giving rise to these clusters. As such, they might not be saturated by any physically-sensible model of impurity--impurity couplings on a given lattice, and so we would actually expect $f(p) \gg m_S(p)$ to hold in real materials. Nevertheless, they represent the limiting scenario for average impurity spin reduction via impurity--impurity couplings (on a given impurity-site lattice and with a given distance-range for the couplings), and if found to be violated, categorically exclude this as a possible explanation for impurity magnetism in the system in question.

\subsubsection{Comparison with experiment}

\begin{figure*}[tbp]
\includegraphics[trim = 0mm 52mm 0mm 0mm, clip, width=1\linewidth]{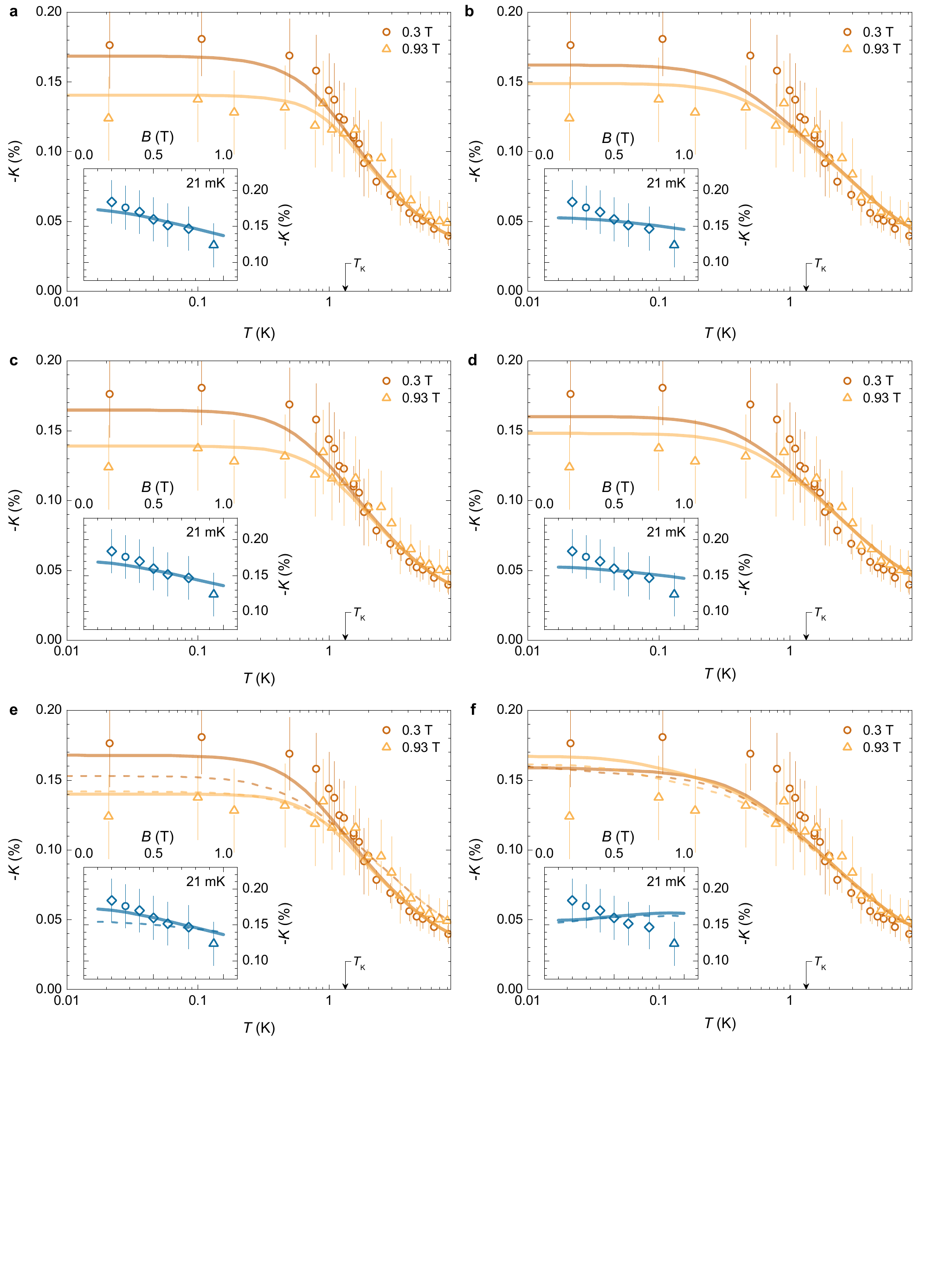}
\caption{
{\bf Supplementary Fig.~2 $\mid$ Muon Knight shift.} 
{\bf a-f}, The Knight shift vs.~temperature (main panels) and magnetic field (insets). The experimental data (symbols) are the same as in Fig.~2a in
the main text while the theoretical curves (lines) are detailed below.
{\bf a}, NRG theoretical prediction computed for
a ``narrow'' conduction band such that $T_K/D \approx 0.5$, where
$D$ is the half-bandwidth of the band.
{\bf b}, The effect of a finite spinon $g$-factor in
a narrow conduction band, $T_K/D \approx 0.2$.
{\bf c}, NRG calculation for a triangular-shaped DOS in a narrow-band case with $T_K/D \approx 0.2$.
{\bf d}, The effect of potential scattering for $V/J=0.5$, with all the other parameters being the same as in the reference model.
{\bf e}, The effect of the XXZ anisotropy in the Kondo exchange
coupling for a dominant longitudinal (Ising) coupling (solid lines) and for a dominant transverse coupling (dashed lines). The magnetic field is oriented along the longitudinal $z$-axis. 
{\bf f}, The effect of a reduced spinon density of states
(DOS) around the spinon chemical potential due to spinon pairing with a hard gap (solid lines) and a BCS-like gap (dashed lines). 
}
\label{figs4}
\end{figure*}

\begin{figure*}[tbp]
\includegraphics[trim = 0mm 0mm 0mm 0mm, clip, width=0.5\linewidth]{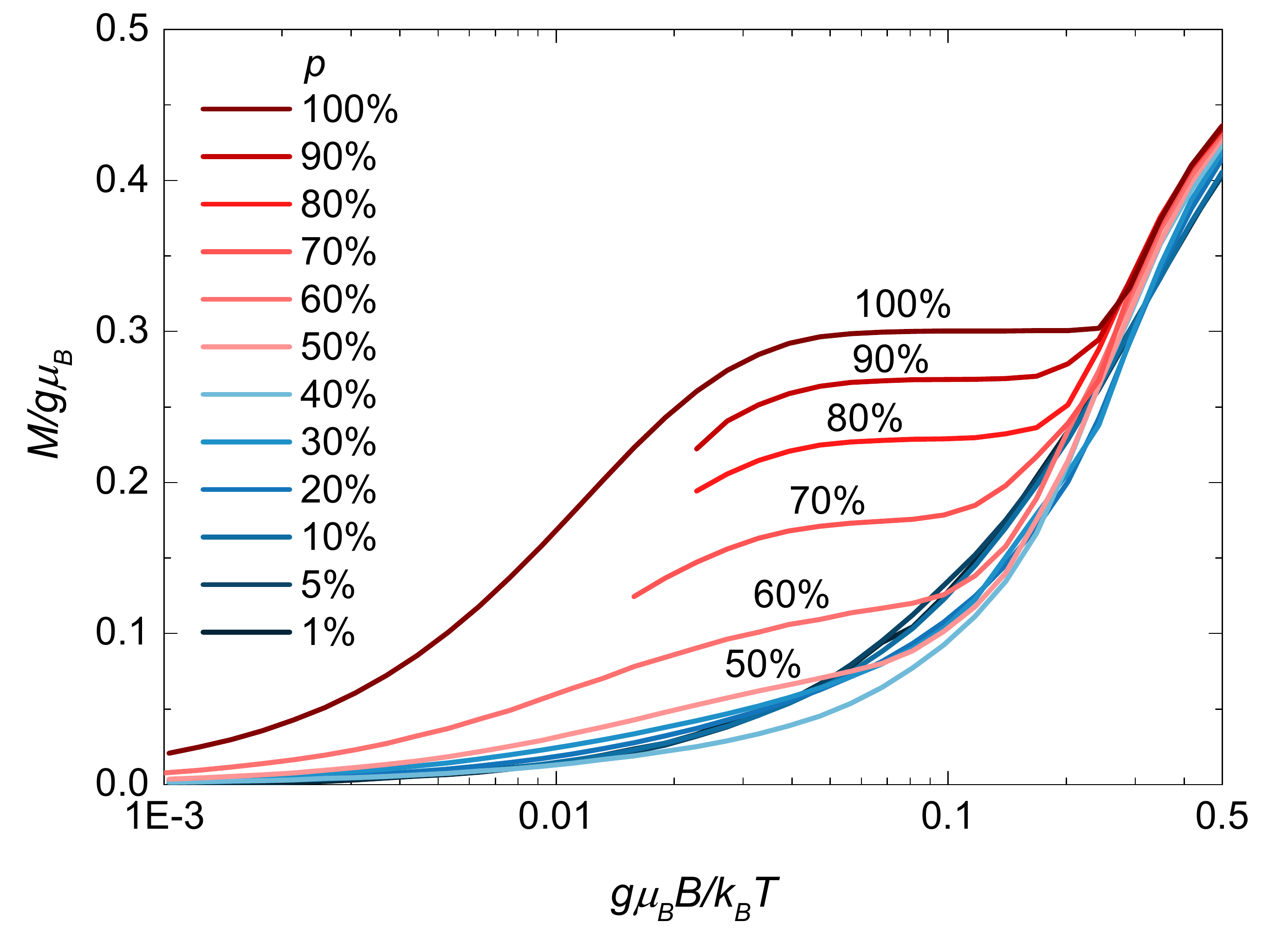}
\caption{ {\bf Supplementary Fig.~3 $\mid$ Magnetization in the Kondo lattice model with a finite
impurity concentration.} The field dependence of the average
impurity magnetization at $T=0.01D$ for a broad range
of impurity concentrations $p$. A magnetization
plateau emerges above $p \sim n$, where $n$ is the electron
density in the lattice; here $n=0.5$. At low impurity concentrations the curves are very similar. Therefore the experimentally-observed $10\%$ impurity concentration in {\bro} can still be considered within the dilute limit.}
\label{figklm}
\end{figure*}

\begin{figure*}[t]
\includegraphics[trim = 0mm 0mm 0mm 0mm, clip, width=1\linewidth]{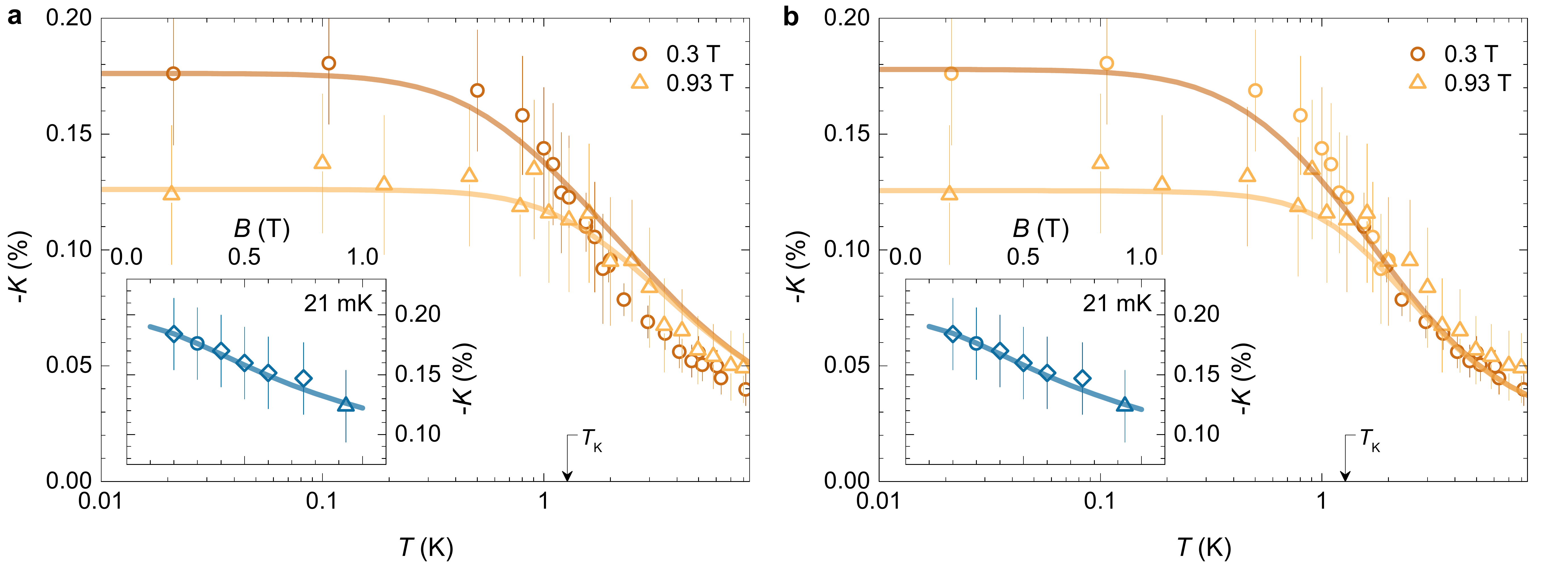}
\caption{
{\bf Supplementary Fig.~4 $\mid$ Muon Knight shift.} 
{\bf a}, Hypothesis of the impurity $g$-factor $g=6.5$ much in excess
of the experimentally determined value of $g=2.3(3)$.
{\bf b}, Hypothesis of a negative bulk $g$-factor, $g_\mathrm{bulk}=-3$.
Here $\rho J=0.5$, so that $T_K/D \approx 0.2$. Better agreement at
low temperatures comes at a price of worse agreement at high
temperatures. This could be improved by taking a smaller value of $J$,
but then $|g_\mathrm{bulk}|$ would need to be even larger.
}
\label{figs6}
\end{figure*}

In {\bro} there are $S = 1/2$ impurities originating from Cu--Zn site-substitution disorder, where the spinful $Cu^{2+}$ ions nominally form a 2D kagome lattice and the spinless Zn$^{2+}$ ions nominally sit at the centres of the kagome-lattice hexagons \cite{li2014gapless}. As the Cu--Zn disorder in {\bro} cannot introduce interlayer couplings between kagome planes, its impurity-site lattice and its couplings are necessarily 2D, in contrast to the 3D impurity-site lattice proposed for the related quantum kagome antiferromagnet herbertsmithite \cite{han2016correlated}. Assuming only nearest-neighbour impurity--impurity couplings, the only two plausible impurity lattices in {\bro} are therefore the triangular lattice (extra Cu$^{2+}$ spin on Zn$^{2+}$ sites) or the kagome lattice (missing Cu$^{2+}$ spin due to extra Zn$^{2+}$). The former has a coordination number of $z = 6$, while the latter has $z = 4$. 
The observed concentration of spin-$1/2$ impurities in {\bro} is relatively low at $p = 6$--$9\%$ \cite{li2014gapless}, while the observed average magnetic moment per impurity $0.19$--$0.28 \mu_B$ (see the main text) corresponds to a strong average spin reduction factor of $f(p) = 0.16$--$0.28$, assuming a full impurity magnetic moment in the range $1.0$--$1.2 \mu_B$ typical for Cu$^{2+}$ ions in an oxygen environment \cite{abragam1961principles}.

To compare this with the lower bound on the spin reduction factor $f(p)$, as given by Eq.~(\ref{clustersEq1}) we performed numerical Monte--Carlo calculation of this lower bound $m(p)$ for a variety of impurity-site lattices with coordination numbers ranging from $z = 2$ to $z = 8$ and a range of $p$ from zero to one (Supplementary Fig.~\ref{clustersFig}). The simulation algorithm was similar to the Hoshen--Kopelman algorithm \cite{hoshen1976percolation} and had a guaranteed space complexity of $O(L)$ for simulating an $L \times L$ patch of a 2D impurity lattice ($L = 4001$ in our case).
Each simulation was run 5 times to estimate the statistical uncertainty of the results. The results conclusively show that the inequality $f(p) \geq m(p)$ of Eq.~(\ref{clustersEq1}) is strongly violated in {\bro} for all simulated lattices.
Furthermore, at a ${>}95\%$ confidence level it must be violated for any lattice with $z \leq 12$, as we observe $f(p) < w_1(p)$ for those $z$, but $w_1(p) \leq m(p)$ for any lattice.
 This result is strong enough to also \textit{a priori} exclude both a kagome impurity lattice with nearest-neighbour (NN) and next-nearest-neighbour (NNN) couplings ($z = 8$), as well as a triangular impurity lattice with  NN and NNN couplings ($z = 12$). 
 We thus conclude that a model with only short-range impurity--impurity couplings between randomly-placed impurities is not compatible with the experimentally determined average impurity moment in \bro.

\begin{figure*}[t]
\includegraphics[trim = 0mm 22mm 0mm 4mm, clip, width=1\linewidth]{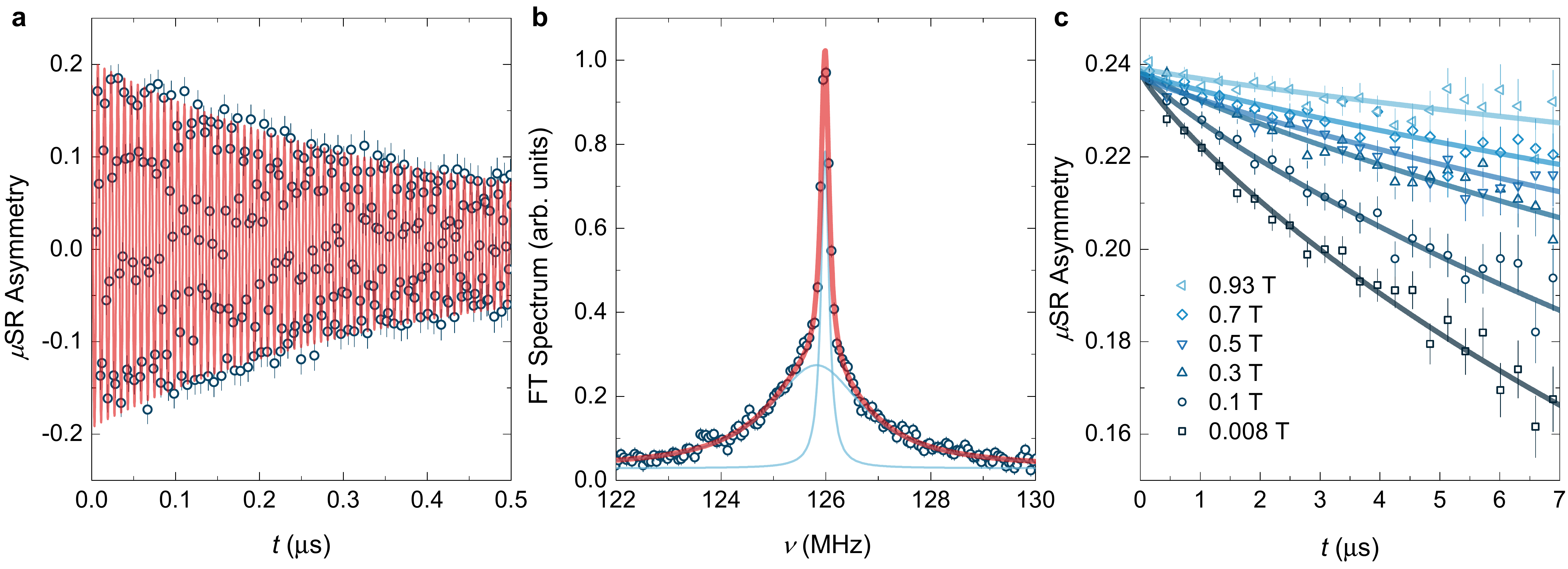}
\caption{
{\bf Supplementary Fig.~5 $\mid$ Muon spin rotation and relaxation measurements.}
{\bf a}, The oscillation of the $\mu$SR asymmetry in \bro, measured in a 0.93~T transverse field at 21~mK (dots). 
The solid line shows a two-component fit with the model given by Eq.~(1) in the main text.
{\bf b}, The corresponding Fourier transform of the $\mu$SR data (dots) and its fit with two Lorentzians (thick line). 
The two components, which are different in width and line position, are shown by thinner lines. 
The narrow (minor) one is the reference signal arising from muons that stop in the silver sample holder, while the broad (dominant) line is assigned to muons stopping in the sample.
{\bf c}, A collection of $\mu$SR asymmetry curves at 0.3~K measured in various longitudinal applied fields (symbols) and fitted with the model given by Eq.~(3) in the main text (lines).}
\label{figs1}
\end{figure*}

\subsection{Perturbations to the Conventional Kondo Model}\label{NRG}

The impurity magnetization $M(B,T)$ of the simple single-channel
$S=1/2$ Kondo model with an isotropic exchange coupling to a wide spinon band
with a constant density of states --- the ``reference model'' --- does not
appear to fully match the measured muon Knight shift (see Fig.~2 in the main text).
Therefore, a number of generalizations of the reference model were considered 
in our attempt to find a better quantitative agreement between
theory and experiment (see Methods in the main text).
These include narrow spinon bands, a finite spinon $g$ factor, energy-dependent bands, potential Kondo scattering, an XXZ anisotropic Kondo coupling, and a band-gap opening.
The effects of these perturbations on the Knight shift are summarized in Supplementary Fig.~\ref{figs4}.

Furthermore, since the impurity concentration in \bro~is 6--9\% \cite{li2014gapless}, additional impurity--impurity interactions might represent important perturbations to the single-impurity Kondo model. 
Their effect on the impurity magnetization for the impurity concentration $p$ spanning 1--100\% is shown in Supplementary Fig.~\ref{figklm}. 

The field dependence of the measured Knight shift at low temperatures can be effectively accounted for by a renormalization of $g$ factors. The effects of renormalizing the impurity and spinon $g$ factors are shown in Supplementary Fig.~\ref{figs6}.

\section{Supplementary Methods}
\subsection{$\mu$SR measurements}\label{muSR}

A typical muon spin rotation dataset measured in a transverse field of 0.93~T at a base temperature of 21~mK is shown in Supplementary Fig.~\ref{figs1}a.
The $\mu$SR asymmetry curve consists of two components (Eq.~(1) in the main text), the dominant one with the amplitude $A_1=0.163(9)$ and the minor one with the amplitude $A_2=0.036(3)$.
Their corresponding frequencies are $\nu_1=125.821(15)$~MHz and $\nu_2=125.977(17)$~MHz.
The dominant component exhibits exponential decay with a decay rate of $\lambda_1=2.2(2)$~$\mu$s, while the relaxation of the minor component is much slower, with $\lambda_2=0.09(3)$~$\mu$s.
The two components are clearly seen in the Fourier transform (FT) spectrum of the same $\mu$SR dataset (Supplementary Fig.~\ref{figs1}b).
The FT spectrum can be fit with two Lorentzian lines with central frequencies $\nu_1$ and $\nu_2$.
The broader, dominant line (in intensity) is notably shifted to lower frequencies with respect to the narrower minor line.
The dominant component is thus assigned to the sample, while the minor component is assigned to the silver sample holder. 
This assignment is further reinforced by the fact that the frequency $\nu_2$ is temperature independent, while the frequency $\nu_1$ changes with temperature, giving a  muon Knight shift that depends on temperature (see Fig.~2a in the main text).

A set of typical muon spin relaxation measurements in various longitudinal applied fields at the temperature of 0.3~K is shown in Supplementary Fig.~\ref{figs1}c.
A stretched-exponential model given by Eq.~(3) in the main text was fit to each dataset.
The initial muon asymmetry was found to be $A_1+A_2=0.239(4)$, while the constant background signal was set to $A_2=0.053(3)$, so that the ration $A_2/A_1$ was the same as the one found in the muon spin rotation measurements.

\end{document}